\documentclass[10pt,letterpaper]{article}

\usepackage{cogsci}
\cogscifinalcopy
\usepackage{pslatex}
\usepackage{apacite}
\usepackage{wrapfig}
\usepackage{graphicx}
\graphicspath{ {./images/} }
\usepackage{soul} 
\usepackage{color}
\usepackage{xcolor,colortbl}
\usepackage[small]{caption}

\title{Fast and flexible: Human program induction in abstract reasoning tasks}

\author{{\large \bf Aysja Johnson$^1$ (aysja.johnson@nyu.edu), Wai Keen Vong$^2$ (waikeen.vong@nyu.edu),}\\{\large \bf Brenden M. Lake$^{1,2}$ (brenden@nyu.edu), \& Todd M. Gureckis$^1$ (todd.gureckis@nyu.edu)} \\
  $^1$Department of Psychology, New York University; 
$^2$Center for Data Science, New York University}
  
\begin{document}

\maketitle

\begin{abstract}

The Abstraction and Reasoning Corpus (ARC) is a challenging program induction dataset that was recently proposed by \citeA{chollet2019measure}. Here, we report the first set of results collected from a behavioral study of humans solving a subset of tasks from ARC (40 out of 1000). Although this subset of tasks contains considerable variation, our results showed that humans were able to infer the underlying program and generate the correct test output for a novel test input example, with an average of 80\% of tasks solved per participant, and with 65\% of tasks being solved by more than 80\% of participants. Additionally, we find interesting patterns of behavioral consistency and variability within the action sequences during the generation process, the natural language descriptions to describe the transformations for each task, and the errors people made. Our findings suggest that people can quickly and reliably determine the relevant features and properties of a task to compose a correct solution. Future modeling work could incorporate these findings, potentially by connecting the natural language descriptions we collected here to the underlying semantics of ARC. 
\end{abstract}

\section{Introduction}

\begin{figure}[t]
    \centering
    \includegraphics[width=0.5\textwidth]{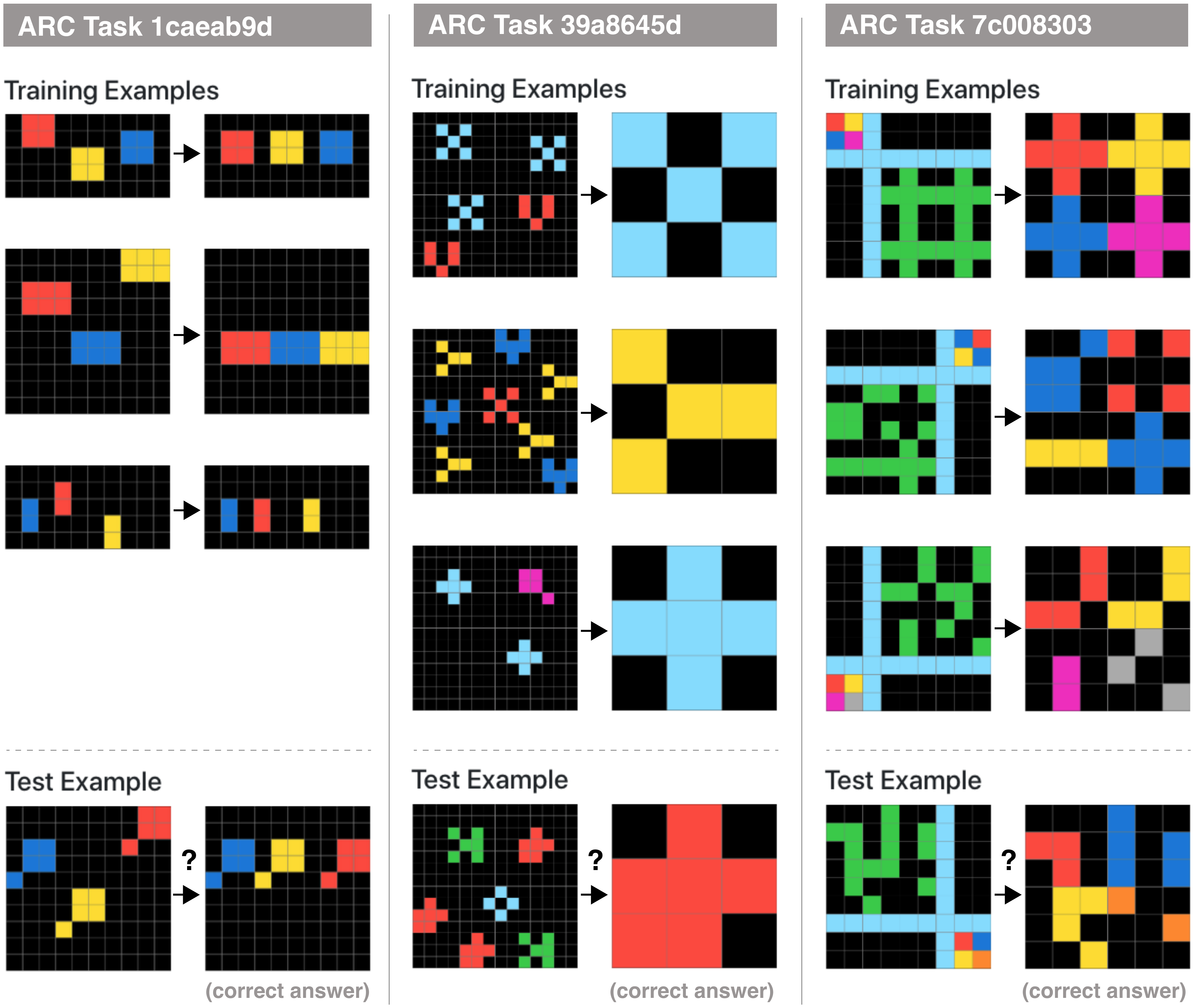}
    \caption{\textbf{Three example ARC tasks}. Each column contains a set of ``training examples'' and a ``test example'' for an ARC problem.  The training examples contains a number of input patterns that point ($\rightarrow$) to the corresponding output.  Here there are always three input$\rightarrow$output pairs but there can be fewer or more.  The rules for transforming the input examples to the output examples are unique for each ARC task. The left task (which we refer to later as the \textbf{box alignment task}) requires aligning the red and yellow objects to the blue object along the vertical axis. The middle task requires counting each object and returning the most common object. The right task requires mapping the four colors in the 2x2 quadrant onto each quadrant of the green object. Understanding of the pattern is assessed with the text example. Here a single test input is provided and the agent has to create the expected output.  The correct answer is displayed here for each problem but the agent does not have access to this.}
    \label{fig:ex_grids}
\end{figure} 

Despite recent advances in AI, contemporary systems still lack the ingenuity and flexibility of human intelligence.  One factor that limits progress in the field of AI is the focus on narrow benchmark challenge problems (e.g., ImageNet \cite{deng2009imagenet}, Arcade Learning Environment \cite{bellemare2013ale}). Although astounding progress has been made on such challenge problems, they often don't require the type of flexible, creative, and abstract reasoning that is the hallmark of human cognition. Put simply, as impressive as performance on ImageNet is, correctly identifying objects in photographs is unlikely to be the central assessment in a meaningful test of human intelligence.  

To address this and other concerns, \citeA{chollet2019measure} proposed a novel type of machine learning challenge problem named the Abstraction and Reasoning Corpus (or ARC). Unlike past machine learning challenge problems, ARC primarily emphasizes a type of perceptually simple, yet conceptually challenging, program induction. Each task involves 2-6 input patterns that are each paired with an output.  The agent is tasked with using this small sample of input-output relationships to infer the underlying program or procedure that modifies the input to create the output. Performance is tested by presenting a novel input pattern, and the agent has to specify what the expected output would be from scratch (see Figure~\ref{fig:ex_grids}).


Several aspects of ARC distinguish it from standard machine learning benchmarks. First, each of the 1000 tasks in the corpus is unique and the various concepts in each rely on many different aspects of knowledge, including: objects, relations, geometry, number, symmetry, and quantification. Second, agents need to manually generate their own response, including specifying the size of the output grid (grids range in size from 1x1 to 30x30), and the colors for each cell (there are 10 possible colors), making the task substantially more open-ended than traditional machine learning challenge problems. Finally, successful solutions to many ARC problems rely on a type of abductive reasoning, often leveraging compositional rules and relationships between identified objects, parts, and reference frames.  As a result, ARC presents a compelling task environment for studying aspects of higher-level cognition and intelligence.

The winner from a recent Kaggle challenge for ARC\footnote{The solution is hosted on their GitHub repository: https://github.com/top-quarks/ARC-solution}, using a program synthesis approach, was only able to solve 21\% of the tasks from the test set, providing a sense of the difficulty for generalization to new tasks with such a limited number of examples. The Kaggle algorithm was built using a domain specific grammar including functions such as `Move', `getSize', and `count', all specified by hand, and the implementation generally lacks cognitive plausibility due to the extensive search requirements for solving each problem.  
 
Given that all of the tasks in ARC were designed by a single, highly sophisticated human (sometimes with the assistance of simple programs), it is unclear whether such low performance by the entrants to the Kaggle challenge would be substantially different from the performance of an average human. In this paper, we report the first behavioral dataset collected on human performance in ARC (to our knowledge), examining performance on a subset consisting of 40 tasks from the ARC training set, with multiple participants per task. Our goal was both to assess how people perform on this benchmark, as well as to understand some of the problem solving mechanisms that they bring to bear on the tasks.

To foreshadow, our results show that humans perform well on ARC---each task we studied was solvable by at least one participant and the average accuracy across all of the tasks was 83.8\%.  In addition, we examined a wide variety of other behaviors generated by people in solving these tasks, ranging from the sequence of actions participants performed to generate the correct output, to natural language descriptions provided about the concepts underlying each task, to the kinds of errors people made. These behavioral phenomena provided us with additional windows to understanding the kinds of inductive biases that contribute to people's success.

\subsection{Program Induction, Abduction, and Probabilistic Language of Thought}

Before describing our experiment, in this section, we consider the ARC challenge in the context of prior work on program induction in cognitive science.


Within cognitive science, much of the work that is most reminiscent of ARC has been in the development of probabilistic language-of-thought (pLOT) models \cite{Piantadosi2011,Goodman2014,piantadosi2016four}. These models assume that hypotheses can be represented as programs specified in a representation language such as first-order logic or lambda calculus. The goal of these kinds of models is to infer the underlying program using Bayesian inference, trading off between the complexity of the program and how well it captures the data. They have been used for modeling a variety of concept learning tasks, ranging from Boolean concepts \cite{goodman2008rational} to complex probabilistic programs \cite{piantadosi2016logical, bramley2018grounding}.

In many experiments testing this framework, the set of primitives and the underlying grammar is itself used to generate the concepts used in experiments. Therefore the set of concepts studied are limited by programs that are easily expressed with a specific grammar. In contrast, the tasks in ARC were hand-designed without reference to a pre-specified grammar. Thus, the types of concepts in ARC are more \textit{abductive} in nature, rather than \textit{inductive}, requiring people to flexibly generate new rules or concepts on the fly, or to determine the relevant variables to bind for a given task. This style of abstract reasoning is one that is less well understood in cognitive science from a computational standpoint, although it is similar to other open-ended puzzles such as Bongard problems where the set of potential rules is also unconstrained \cite{hofstadter1979godel}.


Another aspect of ARC, mentioned above, is that agents are required to generate the correct response from scratch, even specifying the variable size of the output grid.  Few prior studies have looked at people’s ability to generate outputs from novel inputs based on a limited set of training examples (exceptions include  \citeA{mitchell1988copycat}'s work on CopyCat and recent work on compositional induction, \citeNP{lake2019human,rule2018learning}). Additionally, the hope is that tasks with higher-dimensional outputs will reveal more about peoples' mental representations than simpler forced-choice judgments.

\section{Experiment}

\begin{figure}
    \centering
    \includegraphics[width=0.45\textwidth]{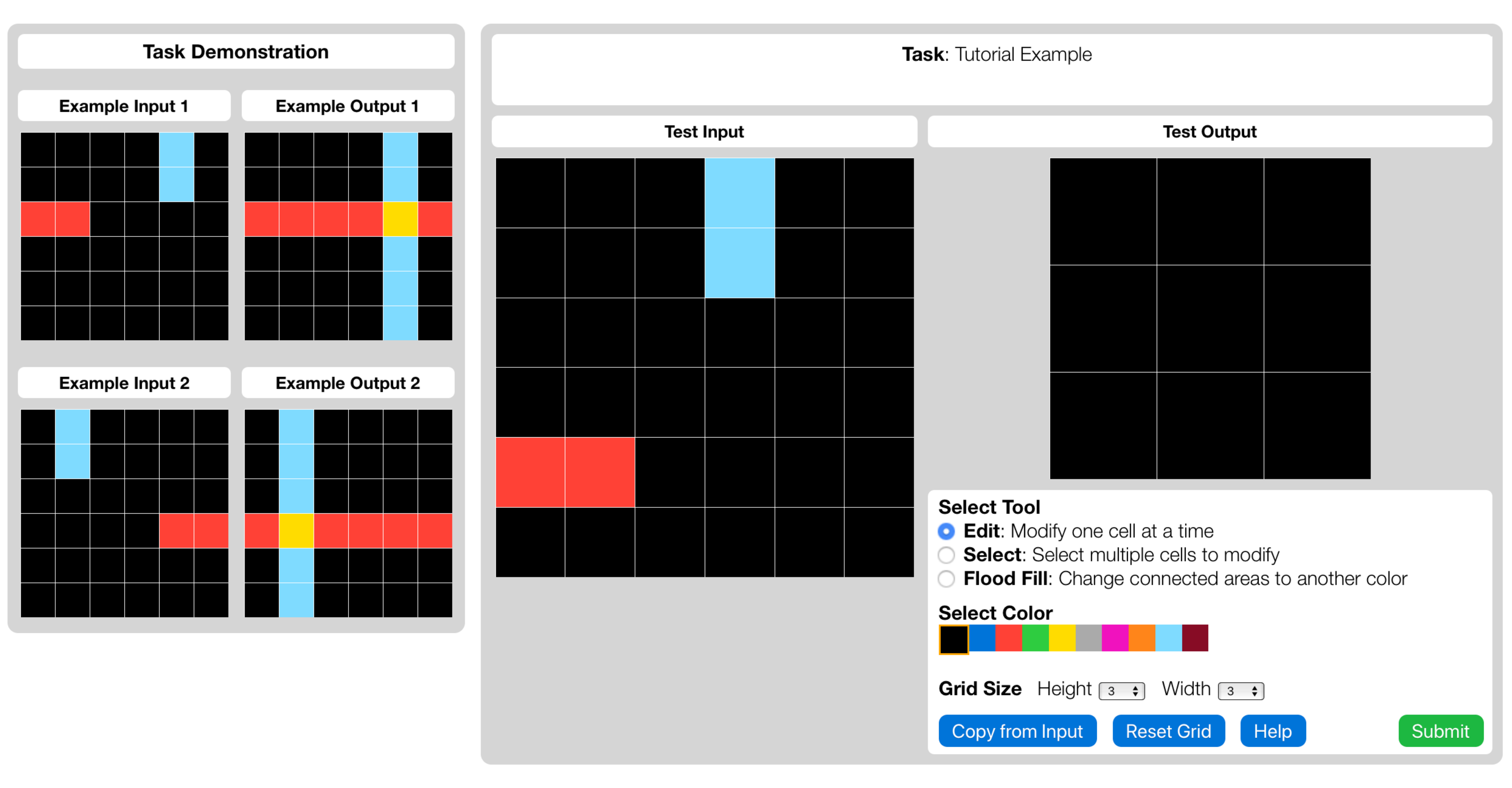}
    \caption{\textbf{ARC User Interface.} On each task, participants were presented with a limited number of example input-output pairs on the left, and their goal was to generate the correct output for a new test input presented on the right. Here we demonstrate the tutorial task presented to participants, where the rule is to extend the red line horizontally, the cyan line vertically, and to color the intersection yellow. The user interface contained a number of different tools for participants to flexibly generate solutions to ARC tasks.}
    \label{fig:arc_interface}
\end{figure}

\subsection{Methods}

\textbf{Participants.}
We recruited 95 participants (57.7\% male, 39.9\% female, 2.4\% other) from Amazon Mechanical Turk using the psiTurk platform \cite{gureckis2016psiturk}. Participants varied in age from 21--70 years (\textit{M} = 39.3, \textit{SD} = 10.3). They were compensated \$7.50 plus a potential one dollar bonus if they succeeded at a randomly selected task with an adequately descriptive written solution (see below).

\textbf{Design.}
Forty tasks were randomly selected from a restricted portion of the training set of the Abstraction and Reasoning Corpus for study, capturing a variety of the kinds of abstract reasoning required to solve these tasks. We restricted ourselves to sampling from tasks which only had a single test item, and where the output grid was no larger than 15x15. Due to time constraints, each participant was randomly assigned to complete 10 out of the 40 selected tasks, resulting in 23.5 participants per task on average.

\textbf{Procedure.}
Participants  were first provided with instructions about the experiment and the user interface, followed by a tutorial task as shown in Figure~\ref{fig:arc_interface}.  To continue, participants had to generate the correct test output for the very simple tutorial task. They were then required to answer three comprehension questions correctly (e.g., ``how many attempts per task will you get?'') in order to advance to the experiment. 

The main experiment consisted of 10 ARC tasks, randomly selected from the set of 40 tasks described above. For each task, participants were presented with between 2 to 6 input-output pairs for training, and a single test input. The goal was to generate the correct test output for each task from scratch, starting from a blank 3x3 grid, using a variety of tools with the built-in editor. 

The overall procedure was designed to reflect the original procedure described in \citeA{chollet2019measure} as closely as possible, in order to allow for human-machine comparisons on ARC.
Participants could edit single grid cells one at a time based on the currently selected color, or edit multiple cells using a selection tool. They could also apply a flood fill operation which colored all neighboring cells of the same color to a different color. Additionally, participants could resize the height and width of the output grid to the desired size, as well as copy the test input to the test output grid.

Participants were allowed three attempts to generate the correct test output, and they were given binary feedback about whether their response was correct or incorrect. Upon a correct submission or three incorrect submissions, participants moved directly onto the next task. 

Additionally, participants were asked to write a description of the solution for transforming input grids to output grids for each task. They were asked to do this once before they received any feedback after their first submission (i.e., after they submitted their response but before they knew whether they were right or wrong). If their first attempt was correct they only submitted one written solution. If it was incorrect, they were asked to submit another written description after they submitted a correct attempt or submitted their third incorrect attempt. Although the tasks were not timed, participants completed the entire experiment in 41 minutes on average (range 17.8 - 72.4).

\section{Results}

\begin{figure}[t]
    \centering
    \includegraphics[width=0.5\textwidth]{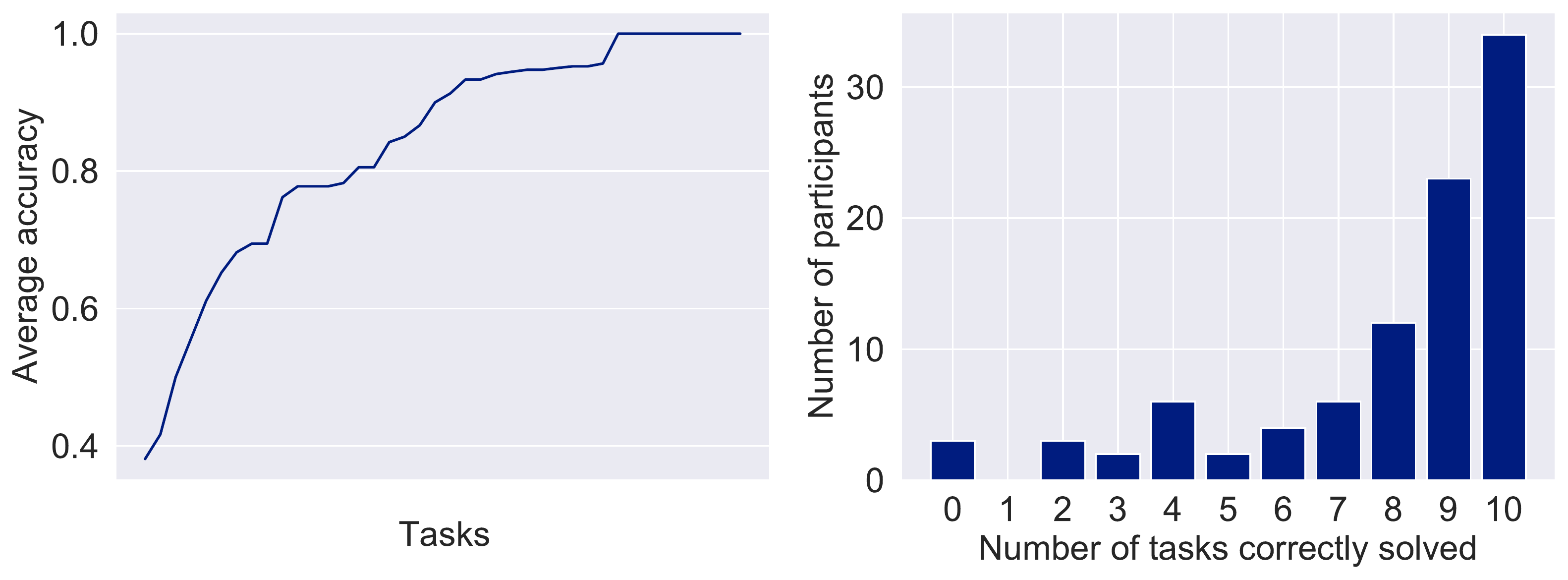}
    \caption{\textbf{Performance on ARC by task and participant}. The left figure shows the average accuracy across all 40 tasks, sorted by accuracy. The figure on the right shows how many tasks each participant solved correctly. Although there is considerable variability at both the task and participant level, performance was generally quite high.}
    \label{fig:accuracy}
\end{figure} 

\begin{figure}[t!]
    \centering
    \includegraphics[width=0.5\textwidth]{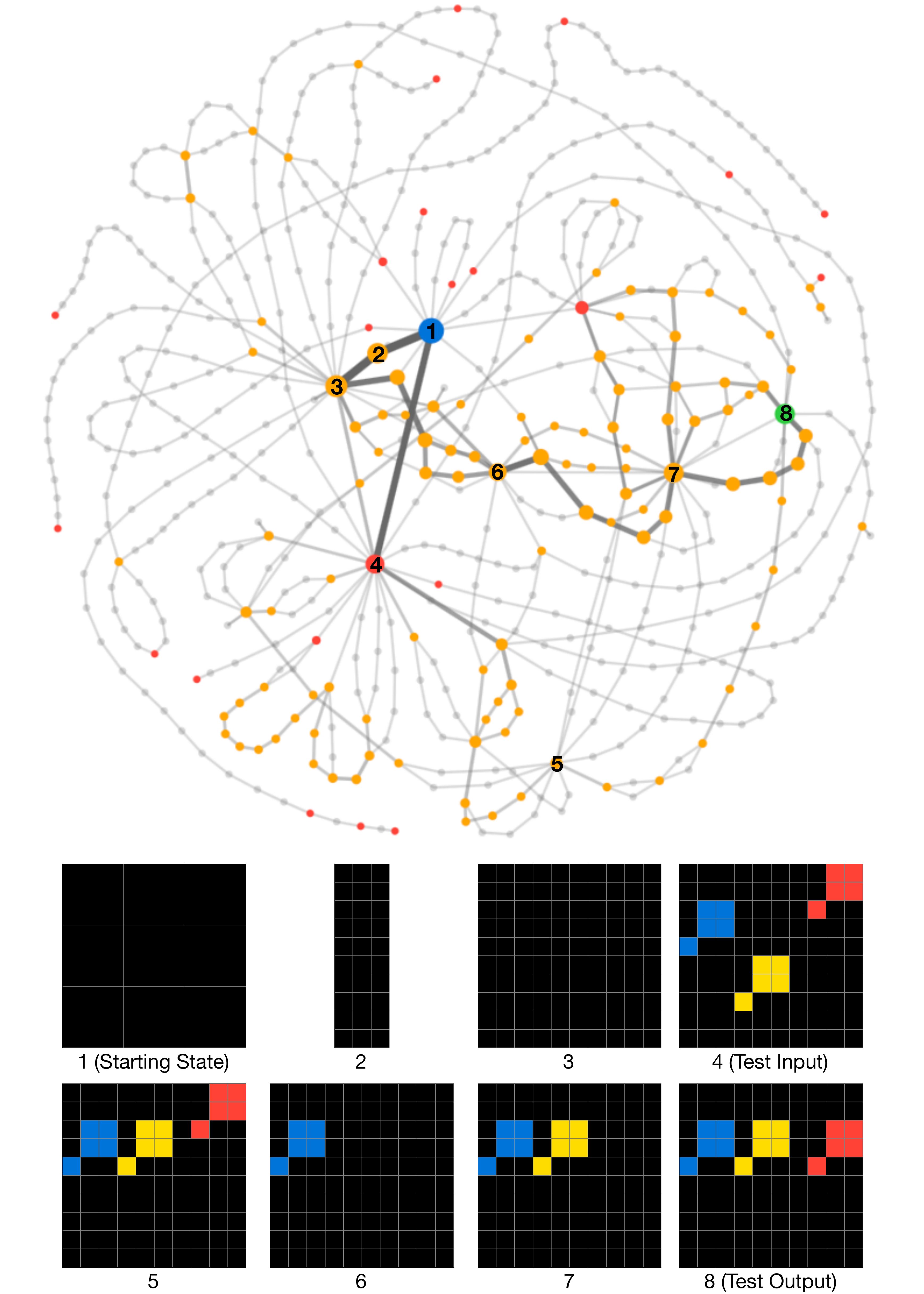}
    \caption{\textbf{State space graph for the box alignment task}. Each node represents an output state and each edge represents an action connecting two states. The colors of the various nodes depict the starting state (blue), the correct output state (green), incorrect submissions (red), and highly visited states (yellow). Larger nodes and thicker edges represent more frequent visits. Additionally, certain bottleneck states shown below the graph highlight that the sub-goals participants used to generate their responses were strongly object-based for this task, by either generating the correct output from scratch one object at a time, or copying the input first and moving the yellow and red objects to their correct location.}
    \label{fig:state_space_graph}
\end{figure}

Visualizations of all participant responses are available at \url{https://arc-visualizations.github.io}.\footnote{The website also contains state space graphs, participant errors, and language descriptions.}
We first analyzed overall performance, using accuracy defined as producing the correct test output for a task within three attempts. Overall, our results show that people performed strongly on the tasks we studied (\textit{M} = 83.8\% per task, \textit{SD} = 16.7\%). In Figure~\ref{fig:accuracy}, we show the distribution of accuracy across tasks, showing considerable variation. While most tasks were easily solvable by the majority of participants (65\% of tasks have 80\% or higher accuracy), only 38.1\% of participants were able to generate the correct output in three attempts for the most challenging task. Qualitatively, tasks that relied on logic, rotations, or flips tended to be the most difficult, while tasks that involved simple color manipulations (such as color inversions or filling objects with colors) were easier. We also observed considerable variation in accuracy across participants as shown in Figure~\ref{fig:accuracy} (\textit{M} = 8.38 of 10 tasks per subject, \textit{SD = 2.7}), however the modal frequency for tasks solved was 10 out of 10.

The average time to complete each task was 3 minutes and 6 seconds (\textit{SD} = 2 minutes and 37 seconds). The average time between first seeing the task and taking the first action was 36 seconds (\textit{SD} = 1 minute and 7 seconds). In addition, the average description length (across incorrect and correct submissions) provided by participants at the end of each task was 20 words (\textit{SD} = 5 words). Participants took 1.59 attempts on average (\textit{SD} = 0.46). 

Although the Kaggle algorithms are not intended as candidates for human cognition, we nevertheless find it informative to note the differences between these algorithms and human performance in order to further elucidate the gap between machine learning systems and human intelligence. The Kaggle algorithm achieved 57.5\% accuracy on the 40 tasks we chose from the ARC training set (relative to 83.8\% for humans). The Spearman rank correlation between human accuracy and the Kaggle algorithm accuracy was 0.35 (\textit{p} $<$ 0.05), although this was driven heavily by the algorithm failing to capture the hardest problems in the set while sometimes capturing the easiest ones.

In order to quantify which factors were linked to solving a given task, we fit a logistic mixed-effects model predicting success on each subject and task pairing using average description length per task as a fixed effect, with random intercepts for participants and tasks. We find a significant, negative effect of average description length on task accuracy (\textit{b} = -0.17, 95\% CI: [-0.315,-0.020], $p = 0.03$). This is in line with recent work  \cite<e.g.>{lupyan2012labeling,lupyan2020does} suggesting that problem difficulty increases as description length increases and as nameability (the ease with which a concept can be named) decreases; and that language biases perceptual processing towards certain concepts. Furthermore, we find a negative correlation (\textit{r}=-0.50) between average description length and average accuracy per task. 

\subsubsection{Action Sequences}

\begin{figure}[t]
    \centering
    \includegraphics[scale=0.4]{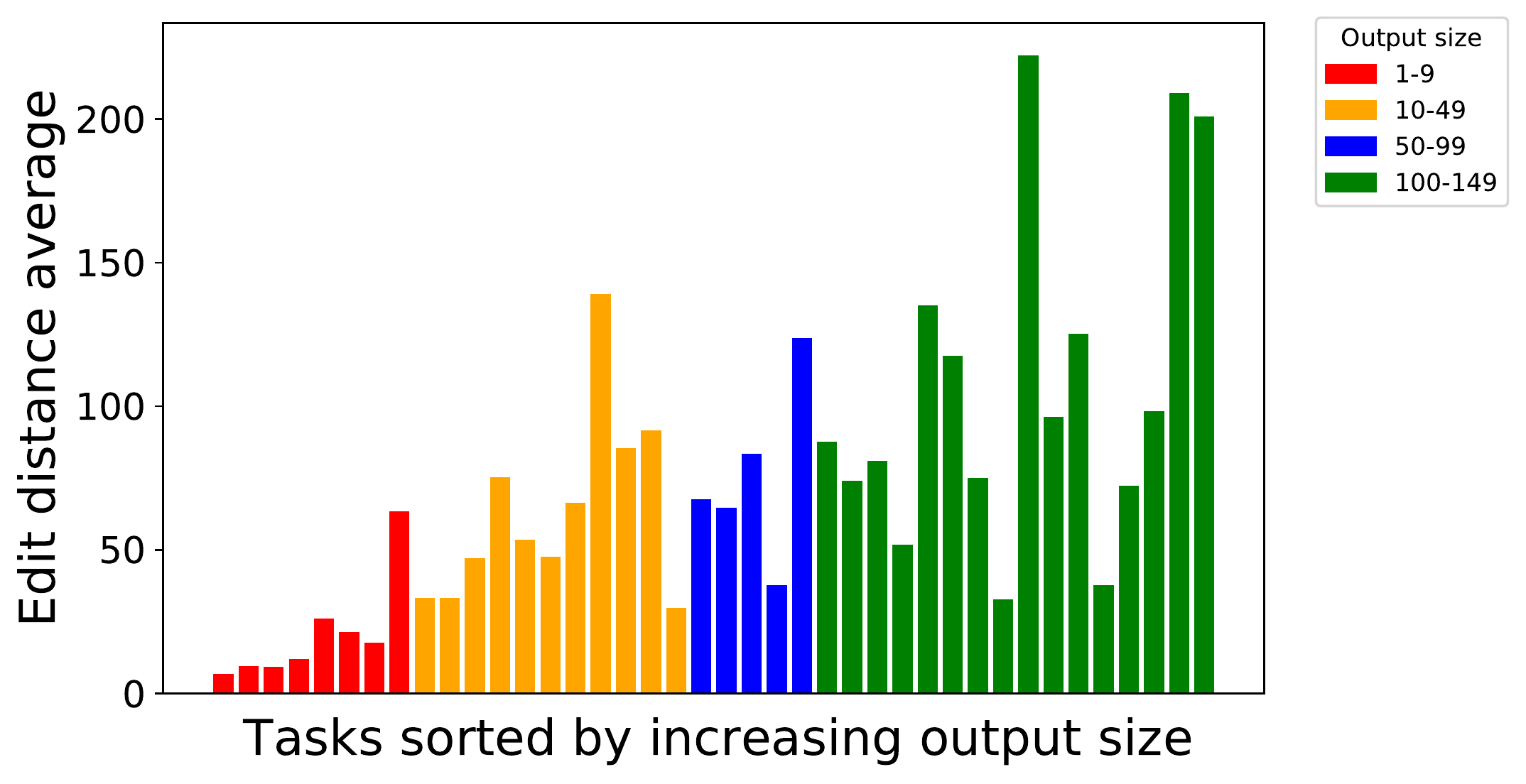}
    \caption{\textbf{Variability in action sequences across tasks}. This figure shows the variability in edit distances for all combinations of action sequences within a task, across all 40 tasks. The tasks are sorted according to the true output grid size (length*width). Edit distance will tend to be higher the wider the range in string length, so we should expect to see overall variance increase with output size. However, within different output grid sizes there is still a range of variability, suggesting that divergence in action sequences is driven by more than output size.}
    \label{fig:edit_dist_fig}
\end{figure}

As participants were required to generate their solutions to ARC tasks, we were able to collect a rich behavioral dataset of the sequences of actions participants performed to generate their responses. Figure~\ref{fig:state_space_graph} displays a state space graph constructed from the action sequences of all of the participants for the box alignment task (referenced in Fig~\ref{fig:ex_grids}) depicting the set of visited output states and transitions to other states based on their actions while generating responses. We found that the majority of participants' first actions are to  either manually resize the output grid to the width and height of the intended solution (Figure~\ref{fig:state_space_graph} node 3), or to copy the test input grid (Figure~\ref{fig:state_space_graph} node 4). For this particular task, we can see that although there is substantial variability in the action sequences participants pursue within this task, participants converge on a few particular paths that pass through some common intermediate bottleneck states. We qualitatively found that the bottleneck states are typically representative of task-relevant objects. Although further research is needed, we think this object-centric action planning reveals an important difference between human and machine solutions to ARC to date.


We also examine the similarity of action sequences between participants across tasks. To do so, we quantify the distance between two action sequences as the Levenshtein edit distance. To perform this analysis, every state which was visited by at least one participant was assigned a number; these numbers were then assembled into a string based on the order of actions for each participant. We then computed the pairwise edit distance for every combination of participants within each task and took the average. Overall, the average edit distance across all tasks (filtered by correct attempts) was 74 edits (\textit{SD} = 53). 
Figure \ref{fig:edit_dist_fig} shows the overall pattern of action sequence similarity across tasks. 
\vspace{5mm}

\subsubsection{Natural Language Descriptions}
In addition to the action sequences, we also analyzed the written natural language descriptions from the end of each task. The descriptions we collected were free form, although participants were encouraged to provide descriptions with enough detail so that they would be useful to future participants to reconstruct the correct test outputs.

\begin{table}[t!]
\begin{tabular}{p{0.08\textwidth}|p{0.37\textwidth}}
\textbf{category}  & \textbf{top unigrams}           \\ \hline
\textbf{color}     & blue (396), color (353), red (244), colors (158)\\ 
\textbf{size}      & size (58), 2x2 (33), 3x3 (33),  4x4 (21)\\
\textbf{location}  & right (122), left (98), bottom (82), where (80)\\
\textbf{relation}  & same (200), match (36), part (22), between (22)\\
\textbf{object} & squares (388), square (221), blocks (124) \\
\textbf{geometric} & line (136), lines (77),  corner (52), diagonal (51)\\
\textbf{number}    & one (139), number (114), 3 (59), two (54)\\
\textbf{abstract}  & tetris (5), paint (4), vessel (2), flower (2)\\
\textbf{transform} & make (105), fill (87), extend (51), copy (49)\\
\end{tabular}
\caption{\textbf{Top unigrams for each category}. Based on the natural language descriptions, we manually grouped words into 9 distinct content classes. The table above shows the top unigrams from each class along with their frequencies in parentheses.}
\label{table:cat_unigrams}
\end{table}

We filtered the dataset to only include the final natural language descriptions participants provided, and to participants who correctly solved the task, as well as removing stop words.\footnote{Participants who didn't solve a task generally used non-informative descriptions like ``I  have no idea'', which was not meaningful for our analyses.} 
We then categorized all of the words from the remaining set of descriptions into 9 distinct classes that captured the different kinds of concepts people used to describe the tasks (see Table \ref{table:cat_unigrams} for a complete list of the categories, as well as the most frequent words used in each category). Figure \ref{fig:cat_hist_fig} shows the proportion of each class across all correct descriptions. Color words were used the most overall as they are relevant in describing almost every task, with object and geometric words second most. Although abstract words are used the least, they provide some of the most interesting abstractions that people relied on to solve ARC tasks. For instance, in the \textbf{box alignment} task from Fig~\ref{fig:ex_grids} a couple of participants referred to the pixel to the left of the boxes as a ``tail'', an inventive mapping of an existing concept to colored squares on an abstract grid, in a manner that many machine learning algorithms lack the capacity to produce. We also found that participants category use varied from task to task, which can be seen in the variability across tasks in Figure \ref{fig:cat_hist_fig}. Once again, we believe these qualitative analyses provide insight into the conceptual primitives that enable humans to far exceed machine performance.

{\begin{figure}[t]
    \centering
    \includegraphics[width=0.48\textwidth]{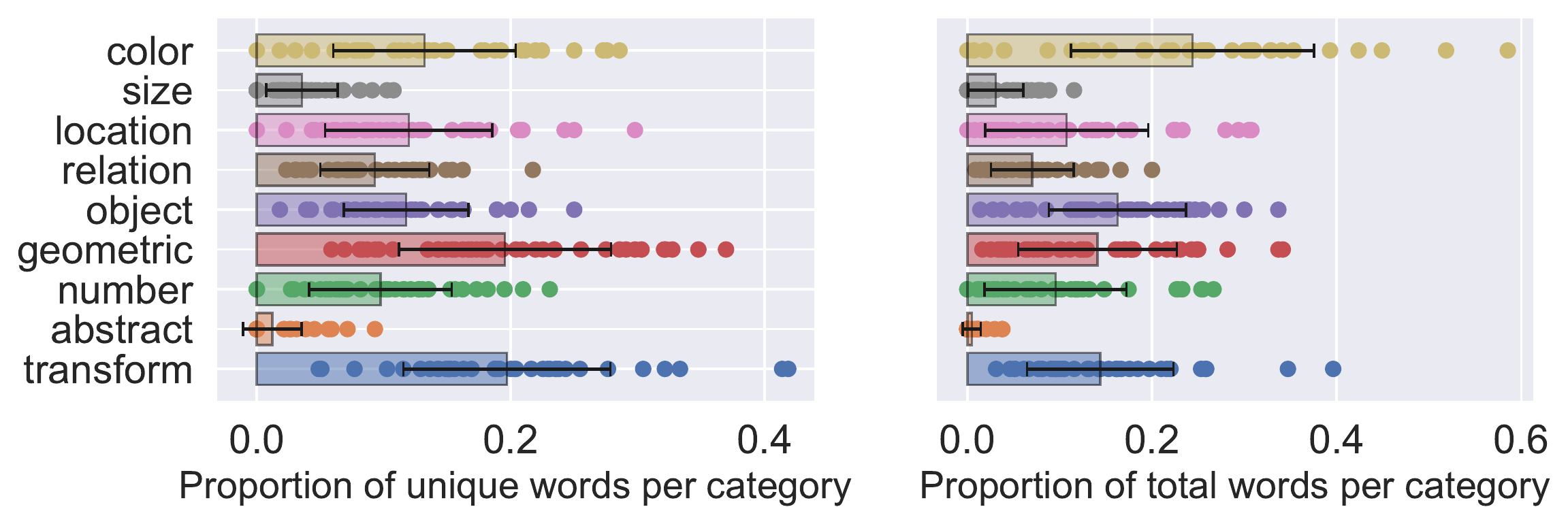}
    \caption{\textbf{Proportion of unique and total words per category}. The left figure shows the normalized proportion of unique words belonging to each category across all descriptions for each task. Geometric and transform words form the largest categories here, implying that people utilize a wider range of these concepts. The right figure shows the normalized proportion of total word counts for each category across all descriptions for all tasks. Here, we find that color words are utilized the most, indicating that people rely on these concepts more overall. Both figures exemplify the variability in concept use across tasks, both in the wide range of word use across tasks (overlaid data points) and the high variance (bars indicate standard deviation).}
    \label{fig:cat_hist_fig}
\end{figure}}



We were also interested in examining how consistent participant descriptions within a task were to one another. In order to measure consistency across descriptions, we use a recently developed measure by \citeA{lupyan2020does} known as naming divergence, which is calculated by (\# of unique words / \# of total words). Naming divergence scores lie between 0 and 1, where a lower naming divergence implies that the set of words used is consistent across participants, whereas a higher naming divergence implies that different participants describe the task differently. We computed naming divergence within tasks and compare this to a shuffled distribution. We generated this by sampling the average naming divergence of shuffled descriptions across tasks (i.e., the average naming divergence of 40 ``tasks'', with randomly sampled descriptions) 1000 times. The shuffled distribution has \textit{M} = 0.68 and \textit{SD} = 0.003, and the true average naming divergence of 0.41 (across tasks) was lower than any of these permutations ($p < 0.001$). This indicates that there was greater consistency in the language used to describe the same task. 


\subsubsection{Errors}

\begin{figure}[t]
    \centering
    \includegraphics[width=0.40\textwidth]{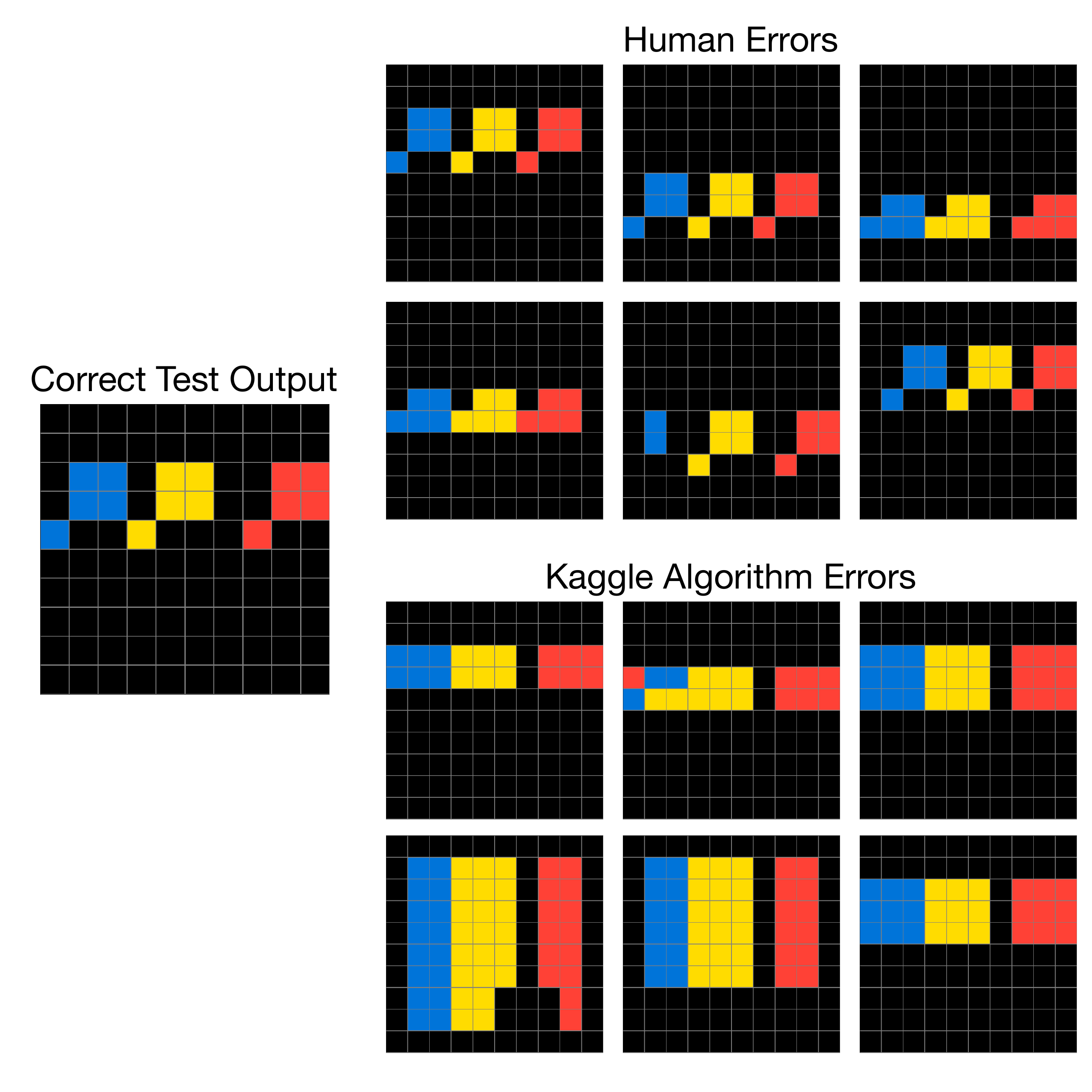}
    \caption{\textbf{Human and Kaggle algorithm errors for the box alignment task}. Here we display the set of most frequent human errors and the full set of Kaggle errors. For the Kaggle competition, each submission was allowed up to three guesses for each test input. Here, two separate submissions are displayed -- each grid is a separate run of the algorithm, i.e., a new search over programs. The human errors obey object priors to a greater extent than the Kaggle errors do.}
    \label{fig:arc_errors}
\end{figure}

Because ARC required participants to generate their responses, the errors people make are informative about the kinds of representations people used to solve these tasks. One of the main aspects we were interested in was whether or not human errors were relevantly close to the correct answer in one or more dimensions. Although it was difficult to formulate measures to quantify this, it was qualitatively apparent that participants do in fact get most of the relevant features of problems right. For instance, Figure \ref{fig:arc_errors} shows how the errors participants made for the box alignment task were overall correct in inferring the right shapes, colors, and one dimension of alignment (along the y-axis). In contrast, although the colors are right for the Kaggle solutions, many of the shapes violate object-like priors from the input grid (e.g., the shapes are egregiously elongated, and one of the shapes appears to wrap around the grid). More examples of errors are available at \url{https://arc-visualizations.github.io}.

\section{Discussion}

We examined human performance on a subset of the Abstraction and Reasoning Corpus, a recently released machine learning benchmark that incorporates elements such as flexible hypothesis generation, few-shot learning, compositionality and program induction. While even the best competitive machine learning systems perform poorly at the task, our results show that humans are very competent in the ARC domain. Given just 2 or 3 input-output examples of a novel concept, the majority of participants were able to apply the correct underlying transformation to create the corresponding test output in three or fewer attempts. Moreover, participants did not need expansive training within ARC, but were able to do so from the basis of a very limited number of tasks (10 at most). The tasks in ARC vary widely in the types of prior knowledge they draw from (e.g. color, relations, objects, transformations, etc.), and yet participants easily recognized and applied the right kinds of knowledge in each task. Additionally, we analyzed data from the action sequences, natural language descriptions, and the errors made on each task. These analyses revealed that there was considerable overlap in how people approached these tasks, and provided additional context and insight into the inductive biases used in ARC. 

There are two ways in which our results challenge existing language-of-thought (LOT) theories of program induction. First, standard LOT-based accounts for concept learning demand a fixed hypothesis space with the primitives defined in advance. Recent work has improved the ways in which these models can generate hypotheses, for example by grounding hypothesis generation based on observations \cite{bramley2018grounding}, or allowing for the LOT to be modified during the learning process \cite{rule2018learning, ellis2020dreamcoder}. However, our analysis of the description words that people used showed that the largest classes of unique words were the geometric and transformation based aspects of ARC tasks. This suggests that set of hypotheses participants are generating in ARC are not based on just a small set of available primitives, but from potentially a much larger class of conceptual background knowledge that might be relevant \cite{murphy1985role}. If so, this poses a challenge for existing LOT-based theories, as it is not clear how the semantics of existing conceptual knowledge could be easily integrated into a minimal LOT. One alternative account that is more parsimonious with our data---especially our findings regarding the relationship between description length and difficulty---is that hypothesis generation uses natural language as a scaffold for generating hypotheses, either instead of or in addition to a symbolic language of thought \cite{carruthers2002cognitive, lupyan2020does, andreas2017learning}.

The second difficulty for existing LOT-based theories to explain ARC is through the lens of object perception. Standard LOT theories parse out the stimuli into symbolic representations that can be easily manipulated. Yet, the notion of what constitutes an object in ARC is flexible due to factors such as occlusion or whether to treat a set of grid cells as a single object or two, matching some of the challenges involved with real-world object perception. Humans can navigate this ambiguity well and can flexibly decide on parses based on the task at hand, a flexibility that is not common in machines.

Overall, this work provides a first step at translating ARC (a machine learning challenge) to a compelling benchmark for program induction in humans as well. Future experimental work on ARC should investigate a wider range of tasks to validate the preliminary results reported here. In addition, once the basic variables are better understood other experimental designs could be overlaid on the task to manipulate aspects of human performance. One interesting observation from the behavioral data was that participants often take up to a minute between starting a new task and performing their first action to generate the output grid, suggesting that a lot of time might be spent thinking about and formulating hypotheses. Another extension of this work would be to develop a computational account for solving ARC tasks that incorporates some of the insights gleaned here about how humans solve these tasks. Our results further suggest that in addition to the set of core knowledge priors (e.g., objects, relations, and counting) described by \citeA{chollet2019measure}, part of the speed and flexibility may also come from incorporating existing conceptual knowledge into the program induction process. Finally, although ARC was designed to push the boundaries of machine intelligence, many of the critiques are also relevant to cognitive science: rich and challenging benchmarks are needed to fully understand and test the limits of broad generalization and abstract reasoning capability in both humans and machines. 

\section{Acknowledgements}
Thanks to Francois Chollet for creating ARC, and for his comments on this work. Additional thanks to Laura Ruis and Yanli Zhou for helpful feedback on the draft. Aysja Johnson and Todd M. Gureckis were supported by the John S. McDonnell Foundation Scholar Award.

\bibliographystyle{apacite}

\setlength{\bibleftmargin}{.125in}
\setlength{\bibindent}{-\bibleftmargin}
\bibliography{refs,brenden_lib}

\end{document}